\documentstyle[12pt,psfig]{article}

\begin{document}

\begin{titlepage}
\begin{center}

{\Large\bf{The Description of $F_2$ at Very High Parton Densities}}
\\[5.0ex]
{\Large\it{M. B. Gay  Ducati $^{1,*}$\footnotetext{$^{*}$E-mail:gay@if.ufrgs.br}}}
 {\it and}
{ \Large \it{ V. P.  Gon\c{c}alves
$^{2,**}$\footnotetext{$^{**}$E-mail:barros@ufpel.tche.br} }}
\\[1.5ex] {\it {$^1$} Instituto de F\'{\i}sica, Univ. Federal do Rio
Grande do Sul}\\ {\it Caixa Postal 15051, 91501-970 Porto Alegre, RS, BRAZIL}\\
 {\it {$^2$} Instituto de F\'{\i}sica e Matem\'atica, Univ. Federal
de Pelotas}\\ {\it Caixa Postal 354, 96010-090 Pelotas, RS, BRAZIL}\\[5.0ex]
\end{center}

{\large \bf Abstract:} The behavior of the structure function at high energies (high
densities) is directly associated to the gluon distribution.  In this paper we analyze
the asymptotic behavior of the structure function at very high densities considering a
leading and a higher twist relation between these two quantities. We verify that  (a)
if the leading twist relation between the structure function and the gluon distribution
is used, we recover the  black disc limit, and (b) a softer behavior is predicted if a
higher twist relation between $F_2$ and $xG$ is considered. While the first behavior is
well-established by the current high density approaches, the higher twist behavior is a
new result. In both cases, the $F_2$ structure function unitarizes and the Froissart
boundary is not violated in the asymptotic regime of high density QCD. \vspace{1.5cm}

{\bf PACS numbers:} 11.80.La; 24.95.+p;

{\bf Key-words:} Small $x$ QCD;   Unitarity corrections; Evolution Equation.

\end{titlepage}

\section{Introduction}

The physics of high-density QCD has become an increasingly active subject of
research, both from experimental and theoretical points of view. In
 deep inelastic scattering (DIS) the parton high density regime 
 corresponds to the
small $x$ region and represents the challenge of studying the interface between the
perturbative and nonperturbative QCD, with the peculiar feature that this transition is
taken in a kinematical region where the strong coupling constant $\alpha _{s}$ is
small. By the domain of perturbative QCD we mean the region where the parton picture
has been developed and the separation between the short and long distance contributions
(the collinear factorization) is made possible by the use of the operator product
expansion (OPE).  The Dokshitzer-Gribov-Lipatov-Altarelli-Parisi (DGLAP) equations
\cite{dglap} are the evolution equations in this kinematical region. These equations
are valid at leading twist, i.e. at a large value of the photon virtuality $Q^2$, where
a  subclass of all possible Feynmann graphs are dominant and the hard coefficient function 
is connected to the proton by only two parton
lines (For more details see e.g. Ref. \cite{cteq}). For small values of $Q^2$, this picture has corrections predicted by  the OPE 
that 
contribute at relative order ${\cal{O}}(1/Q^2)$ and beyond
$[{\cal{O}}(\frac{1}{Q^2})^n, \,n=2,3,...]$. These are commonly called higher twist
corrections.

In the limit of small values of $x$ $(<10^{-2})$, on the other hand, one expects to see
new features inside the nucleon: the density of gluons and quarks become very high and
an associated new dynamical effect  is expected to stop the
further growth of the structure functions. In particular, for a fixed hard scale $Q^2
\gg \Lambda_{QCD}^2$, the OPE eventually breaks down at sufficiently small $x$
\cite{mueplb}. Ultimately, the physics in the region of high parton densities will be
described by nonperturbative methods, which is still waiting for a satisfactory
solution in QCD. However, the transition from the moderate $x$ region towards the small
$x$ limit may possibly be accessible in perturbation theory, and, hence, allows us
to test the ideas about the onset of nonperturbative dynamics.

At this moment, there are several approaches in the literature that propose distinct
evolution equations for the description of the gluon distribution in high density limit
\cite{100,ayala1} \cite{jamal,kov}. In general these evolution equations resum a class
of higher twist diagrams dominant in this kinematical region that contain powers of the
function $\kappa (x,Q^{2})\equiv \frac{3\pi ^{2}\alpha
_{s}A}{2Q^{2}}\frac{xg(x,Q^{2})}{\pi R_{A}^{2}}$, which represents the probability of
gluon-gluon interaction inside the parton cascade. Moreover, these equations match (a)
the DLA limit of the DGLAP evolution equation in the limit of low parton densities
$(\kappa \rightarrow 0)$; (b)the GLR equation and the Glauber-Mueller formula as first
terms of the high density effects. The main differences between these approaches occurs
in the limit of very large densities, where all powers of $\kappa$ should be resummed.
Although the complete demonstration of the equivalence between these formulations in
the region of large $\kappa $ is still an open question, some steps in this direction
were given recently \cite{npbvic,kovner}.

Our goal in this paper is not to demonstrate the equivalence between the distinct
approaches, but to analyze the relation between the gluon distribution $xG$ and the
$F_2$ structure function in the asymptotic regime of very high densities (energies). 
In
general, a leading twist relation is used to relate these quantities  
and the solutions of the evolution equations used as input in the calculations.
 However, as discussed above, the high density approaches resum a class of
higher twist diagrams, which implies that we should careful in our predictions for the
quantities which would be measured. Here we study the behavior of $F_2$,  predicted by a
leading and a higher twist relation, at large densities, where analytical calculations
are possible. In this regime we expect the blackness of the cross section, as predicted
by Gribov many years ago \cite{gribov}. We verify that this behavior is recovered by
the current high density approaches, if a leading twist relation between $F_2$ and $xG$
is used. However, a softer behavior is obtained when we consider that this relation is
modified by the higher twist terms associated to the high density, demonstrating the
importance of this  contribution. In both cases, the Froissart boundary is not
violated.

The paper is organized as follows: in the next section we briefly review the
derivation of the leading and higher twist relations between $F_2$ and $xG$,
as well as the black disc limit. After, in section 3, we derive the
asymptotic solution of the AGL equation at fixed and running $\alpha_s$ and
compare this solution with the predictions of the McLerran - Venugopalan  high density approach.
Finally, in section 4, we present our results and conclusions.

\section{The $F_2$ Structure Function and the Black Disc Limit}

We start from the space-time picture of the $ep$ processes \cite{gribov}. The deep
inelastic scattering $ep \rightarrow e + X$ is characterized by a large electron energy
loss $\nu$ (in the target rest frame) and an invariant momentum transfer $q^2 \equiv -
Q^2$ between the incoming and outgoing electron such that $x = Q^2/2m_N \nu$ is fixed.
In terms of Fock states we then view the $ep$ scattering as follows: the electron emits
a photon ($|e> \rightarrow |e\gamma>$) with $E_{\gamma} = \nu$ and $p_{t \, \gamma}^2
\approx Q^2$, after the photon splits into a $q \overline{q}$ ($|e\gamma> \rightarrow
|e q\overline{q}>$) and typically travels a distance $l_c \approx 1/m_N x$, referred as
the coherence length, before interacting in the nucleon. For small $x$ (large $s$,
where $\sqrt{s}$ is $\gamma^*p$ center-of-mass energy), the photon converts to a quark
pair at a large distance before it interacts to the target.
Consequently, the space-time picture of the DIS in the target rest frame can be viewed
as the decay of the virtual photon at high energy (small $x$) into a quark-antiquark
pair long before the interaction with the target. The $q \overline{q}$ pair
subsequently interacts with the target. In the small $x$ region, where $x \ll
\frac{1}{2mR}$, the $q\overline{q}$ pair crosses the target with fixed transverse
distance $r_t$ between the quarks. Following Gribov \cite{gribov}, we may write a
double dispersion relation for the forward $\gamma^*p$ elastic amplitude  $A$, related
to the total cross section by the optical theorem ($Im \, A = s \sigma(s,Q^2)$), and
obtain for fixed $s$
\begin{eqnarray}
\sigma(s,Q^2) = \sum_q \int \frac{dM^2}{M^2 +Q^2}\frac{dM^{\prime 2}}{M^{\prime 2}
+Q^2} \rho(s,M^2,M^{\prime 2}) \frac{1}{s} \, Im \, A_{q\overline{q} +
p}(s,M^2,M^{\prime 2}) \,\,, \label{sigdis}
\end{eqnarray}
where $M$ and $M^{\prime}$ are the invariant masses of the incoming and outgoing
$q\overline{q}$ pair. If we assume that the  forward $q\overline{q} + p$ scattering does not
change the momentum of the quarks then $A_{q\overline{q} + p}$ is proportional to
$\delta(M^2 - M^{\prime 2})$, and (\ref{sigdis}) becomes
\begin{eqnarray}
\sigma(s,Q^2) = \sum_q \int \frac{dM^2}{(M^2 +Q^2)^2} \rho(s,M^2) \sigma_{q\overline{q}
+ p}(s,M^2) \,\,, \label{sigdis2}
\end{eqnarray}
where the spectral function $\rho(s,M^2)$ is the density of $q\overline{q}$ states,
which may be expressed in terms of the $\gamma^* \rightarrow q\overline{q}$ matrix
element \cite{ryskin}. Using that $M^2 = (k_t^2 + m_q^2)/[z(1-z)]$, where $k_t$ and $z$
are the transverse and longitudinal momentum components  of the  quark with mass $m_q$,
we can express the integral over the mass $M$ of the $q\overline{q}$ in terms of a
two-dimensional integral over $z$ and $k_t$. Instead of $k_t$, it  is useful to work with the
transverse coordinate $r_t$ (impact parameter representation), which is the variable
Fourier conjugate to $k_t$, resulting \cite{nik}
\begin{eqnarray}
F_2(x,Q^2)  =  \frac{Q^2}{4 \pi \alpha_{em}} \sigma (s,Q^2)  = 
\frac{Q^2}{4 \pi \alpha_{em}} \int dz \int \frac{d^2r_t}{\pi} |\Psi(z,r_t)|^2 \,
\sigma_{q\overline{q}}(z,r_t)\,\,, \label{f2target}
\end{eqnarray}
where
\begin{eqnarray}
|\Psi(z,r_t)|^2 = \frac{6 \alpha_{em}}{(2 \pi)^2} \sum^{n_f}_i e_i^2 \{[z^2 + (1-z)^2]
\epsilon^2\, K_1(\epsilon r_t)^2 + m_i^2\, K_0(\epsilon r_t)^2\}\,\,.  \label{wave}
\end{eqnarray}
The photon wave function $\Psi(z,r_t)$ is simply the Fourier transform of the matrix
element for the transition $\gamma^* \rightarrow q\overline{q}$. Moreover,
$\alpha_{em}$ is the electromagnetic coupling constant, $\epsilon^2 = z(1-z)Q^2 +
m_i^2$, $m_i$ is the quark mass, $n_f$ is the number of active flavors, $e_f^2$ is the
square of the parton charge (in units of $e$), $K_{0,1}$ are the modified Bessel
functions and $z$ is the fraction of the photon's light-cone momentum carried by one of
the quarks of the pair. In the leading log$(1/x)$ approximation we can neglect the
change of $z$ during the interaction and describe the cross section
$\sigma^{q\overline{q} }(z,4/r_t^2)$ as a function of the variable $x$. Considering
only light quarks ($i=u,\,d,\,s$) $F_2$ can be expressed by \cite{plb}
\begin{eqnarray}
F_2(x,Q^2) = \frac{1}{4 \pi^3} \sum_{u,d,s} e_i^2 \int_{\frac{1} {Q^2}}^{
\frac{1}{Q_0^2}} \frac{ d^2r_t}{\pi r_t^4}\,\sigma_{q\overline{q}}(x,r_t) \,\,.
\label{f2sim}
\end{eqnarray}
Using that $\sigma_{q\overline{q}} = \frac{C_F}{C_A} (3 \alpha_s(\frac{4}{r_t^2}
)/4)\,\pi^2\,r_t^2\, xG(x,\frac{4}{r_t^2})$ \cite{plb}, where $xG(x,\frac{4}{r_t^2})$ is the
nucleon gluon distribution, we get
\begin{eqnarray}
F_2(x,Q^2) = \frac{2 \alpha_s}{9 \pi} \int_{Q_0^2}^{Q^2} \frac{d Q^2}{Q^2}
\, xG(x,Q^2) \,\,.  \label{f2vio}
\end{eqnarray}
The above expression  explicit the direct dependence of the  $F_2$ structure function
with the behavior of the gluon distribution, which is related to the QCD dynamics at
high energies. In the linear regime, where the high densities can be disregarded, the
small $x$ behavior of the gluon distribution is given by the solutions of the  DGLAP \cite{dglap}
and/or BFKL \cite{bfkl} evolution equations. The common feature of these equations is the steep
increase of $xG$ as $x$ decreases. This steep increase cannot persist down to arbitrary
low values of $x$ since it violates a fundamental principle of quantum theory, {\it
i.e.} the unitarity. In the context of relativistic quantum field theory of the strong
interactions, unitarity implies that the cross section cannot increase with increasing
energy $s$ above $log^2 \,s$: the Froissart's theorem \cite{froi}. In the next section
we will consider the modifications in the behavior of the gluon distribution associated
to the high density effects. The presence of a high parton density implies a slow
growth of $xG$ at high energies, and consequently from Eq. (\ref{f2vio}), of the
structure function. However, the Eq. (\ref{f2vio}) is a leading twist relation which
will eventually breaks down when we consider the higher twist terms in the evolution.

We now consider that the relation between the structure function and the gluon
distribution is modified by the particular type of higher twist terms associated with
the high density corrections (See also \cite{mv2} for a similar calculation in the
infinite momentum frame). In general, the higher twist  contributions  should be
significant at small $x$ and $Q^2$, implying large perturbative corrections to the
conventional leading twist relations in this kinematical region.
 We estimated these corrections considering the $s$-channel
unitarity constraint in the interaction cross section of the quark-antiquark pair with
the target \cite{ayala2}. In this case  the structure function is given by
\begin{eqnarray}
F_2(x,Q^2) = \frac{1}{2\pi^3} \sum_{f=u,d,s} e_f^2 \int_{\frac{1}{Q^2}}^{\frac{%
1}{Q_0^2}} \frac{d^2r_t}{\pi r_t^4} \int d^2b_t \{1 - e^{-\frac{1}{2}%
\sigma^{q\overline{q}}(x,4/r_t^2)S(b_t)}\}\,\,.  \label{f2eik}
\end{eqnarray}
A similar expression was used in Ref. \cite{golec} for a phenomenological analysis of
the $ep$ process, where the geometrical structure of the collision (the
$\vec{b}_t$ dependence) was disregarded and  $xg\propto x^{-\lambda }\,\,(\lambda >0)$ was assumed,
resulting a very good description of the HERA data. The main point of the above
expression is that this resums a large class of higher twist terms, as demonstrated  in
Ref. \cite{bargol}.

The use of a Gaussian parameterization for the nucleon profile function $ S(b_t) =
\frac{1}{\pi R^2} e^{-\frac{b^2}{R^2}}$, where $R$ is the spatial gluon distribution
inside the proton \cite{raio}, simplifies the calculations. We obtain that the $F_2$
structure function can be written as \cite{ayala2}
\begin{eqnarray}
F_2(x,Q^2) = \frac{R^2}{2\pi^2} \sum_{u,d,s} \epsilon_i^2 \int_{\frac{1}{Q^2}%
}^{\frac{1}{Q_0^2}} \frac{d^2r_t}{\pi r_t^4} \{C + ln(\kappa_q(x, r_t^2)) +
E_1(\kappa_q(x, r_t^2))\}\,\,,  \label{diseik2}
\end{eqnarray}
where $\kappa_q = 4/9 \kappa_G = (2 \alpha_s/3R^2)\,\pi\,r_t^2\,  xG_N(x,
\frac{1}{r_t^2})$. This equation allows to estimate the high density corrections to the
structure function in the DLA limit.  Expanding the equation (\ref{diseik2}) for small
$\kappa_q$, the first term (Born term) will correspond to the usual DGLAP equation in
the small $x$ region [Eq. (\ref{f2vio})].

Before we discuss the dynamics for the gluon distribution at high densities, it is
important to consider a general property of the structure function (i.e. the cross
section): the black disc limit. As shown before, when the coherence length $l_c$
considerably exceeds the diameter of a target which is at rest, the virtual photon
transforms into hadron components well before the target, which implies that the small
$x$ physics probes the interaction of various hadron wave packets with a target. The
geometrical limit for the cross section of virtual photon scattering off a nucleon
target follows from the assumption that a target is black for the dominant hadron
components in the wave function of the virtual photon. An identical  assumption is
often used to deduct the Froissart limit  of hadron-hadron interactions at high
energies.

Under the assumption that the interaction is black, we can estimate the total cross
section (\ref{sigdis2}) using the following  approximations: $\rho(s,M^2) \propto M^2$
and $\sigma_{q\overline{q} + p}(s,M^2) = \pi R^2$. Consequently, we can get
$\sigma(s,M^2) \propto R^2 \,ln [(M^2_{max} + Q^2)/(M^2_{min} + Q^2)]$. Using that
$M^2_{min} \approx 4 m_{pi}^2$ whereas $M^2_{max} \propto s$, results that
\begin{eqnarray}
F_2(x,Q^2) \propto Q^2 R^2 \, ln \frac{1}{x} \,\,,\label{f2black}
\end{eqnarray}
which is the black disc limit for the $F_2$ structure function. This bound, derived
from a geometrical analysis, represents the maximum value allowed for the cross
sections at high energies (high densities).  In the next sections we will analyze the
predictions of the high density approaches for the asymptotic behavior of $F_2$ and
verify if this bound is saturated or a softer behavior is predicted.

\section{The AGL High Density Approach}

About seventeen years ago, Gribov, Levin, and Ryskin (GLR) \cite{100} performed a
detailed study of the high density limit of QCD in the double logarithmic approximation
(DLA). They argued that  the physical processes of interaction and recombination of
partons become important in the parton cascade at a large value of the parton density,
and that these high density corrections could be expressed in a new evolution equation
- the GLR equation.  Some years ago  an eikonal approach to  the high density
corrections was proposed in the literature by  Ayala, Gay Ducati, and
Levin (AGL)  \cite{ayala1,ayala2} which improves the GLR approach.  This approach 
address the high density regime from the region
where perturbative QCD is valid by summing corrections to the linear DGLAP evolution
equation.

As we are only interested in the solutions of the AGL equation, we refer the original
papers \cite{ayala1,ayala2} for details in its derivation. Here we only present the AGL
equation, which is given by 
\begin{eqnarray}
\frac{\partial^2 xG(x,Q^2)}{\partial y \partial \epsilon} = \frac{2\,Q^2 R^2%
}{\pi^2} \{ C + ln [\kappa_G (x,Q^2)] + E_1 [\kappa_G (x,Q^2)]\} \,\,, \label{agl2}
\end{eqnarray}
where $C$ is the Euler constant, $E_1$ is the exponential function and the function
$\kappa_G$ is defined by
\begin{eqnarray}
\kappa_G (x,Q^2) \equiv \frac{\alpha_s N_c \pi }{2 Q^2 R^2} xG(x,Q^2)\,\,,
\label{kapag}
\end{eqnarray}
and represents the probability of gluon-gluon interaction inside the parton cascade.
Using the above definition for $\kappa_G$ we can rewrite the expression (\ref{agl2}) in
a more convenient form (for fixed $\alpha_s$)
\begin{eqnarray}
\frac{\partial^2 k_G (y,\epsilon)}{\partial y \partial \epsilon} + \frac{%
\partial k_G (y,\epsilon)}{\partial y} & = & \frac{ \alpha_s N_c}{\pi} \{ C
+ ln [\kappa_G (x,Q^2)] + E_1 [\kappa_G (x,Q^2)]\}  \nonumber \\ & \equiv & F(\kappa_G)
\,\,.  \label{agl3}
\end{eqnarray}

Analyzing the structure of the Eq. (\ref{agl3}) we see that it has a solution which
depends only on $y$. In \cite{ayala1} it was shown that this solution is the asymptotic
solution of the AGL equation. In this case we have that 
 at large values of densities  the asymptotic
solution is given by
\begin{eqnarray}
k_G^{asymp} (y) = \frac{\alpha_s N_c}{\pi} \,\, y \,\,.  \label{kasy}
\end{eqnarray}
Substituting the definition of $\kappa_G$ [Eq. (\ref{kapag})] in the above
solution,  in the  asymptotic regime the behavior of the gluon
distribution is given by (at fixed $\alpha_s$)
\begin{eqnarray}
xG(x,Q^2) = \frac{2N_c Q^2 R^2}{3 \pi^2} \,\, ln \,(\frac{1}{x}) \,\,.
\label{gluonasy}
\end{eqnarray}
Therefore, the gluon distribution does not saturate at small values of $x$,
but is linearly proportional to $ln \, s$ $(s \approx 1/x)$. However, this
behavior is softer than predicted by the DGLAP equation ($xG \propto exp [%
\sqrt{ ln 1/x}]$) and the BFKL equation ($xG \propto x^{- \lambda}$, $%
\lambda > 0$). We obtain  that the gluon distribution presents a partial saturation in its
behavior.

The analyzes above was made at leading order, considering a fixed coupling constant
$\alpha_s$. One question is the possible modifications in the asymptotic behavior
associated to next-to-leading order corrections as, for instance, the running of
$\alpha_s$. In order to answer this question we need to extend the Refs.
\cite{ayala1,ayala2} for  running coupling constant, as we will do in the following.

 Since the QCD coupling constant
is given by $\alpha_s (Q^2) = 4 \pi /(\beta_0 \epsilon)$, where $\beta_0 = 11 - 2/3
n_f$ ($n_f$ is the number of flavors) and $\epsilon = ln \, Q^2/\Lambda_{QCD}^2$, the
relation between the gluon distribution and the function $\kappa_G$ can be expressed by
\begin{eqnarray}
xG(x,Q^2) = \frac{2 Q^2 R^2}{N_c \pi} \frac{\beta_0}{4 \pi}\, \epsilon \,
\kappa_G (x,Q^2) \,\,.
\end{eqnarray}
Using this result in Eq. (\ref{agl2}) we obtain that the AGL equation with
running $\alpha_s$ is given by
\begin{eqnarray}
\frac{\partial^2 k_G (y,\epsilon)}{\partial y \partial \epsilon} + \left(
\frac{1}{\epsilon} + 1 \right) \frac{\partial k_G (y,\epsilon)}{\partial y}
& = & \frac{N_c \alpha_s (Q^2)}{\pi} \{ C + ln [\kappa_G (x,Q^2)]  \nonumber
\\
& + & E_1 [\kappa_G (x,Q^2)]\}  \nonumber \\
& \equiv & H(\kappa_G) \,\,.  \label{runagl}
\end{eqnarray}
The AGL equation at fixed $\alpha_s$ is a direct consequence of the above
equation in the limit of large $\epsilon$. Moreover, this equation also has
a solution which depends only on $y$. In this case we have
\begin{eqnarray}
\frac{\partial k_G^{asymp} (y,\epsilon)}{\partial y} = \frac{\epsilon}{1+
\epsilon} H(\kappa_G)  \label{asyrun}
\end{eqnarray}
with the solution
\begin{eqnarray}
\int_{ k_G^{asymp} (y = y_0)}^{ k_G^{asymp} (y)} \frac{ d \kappa_G^{\prime}}{%
H(\kappa_G^{\prime})} = \frac{\epsilon}{1+ \epsilon} \, (y - y_0) \,\,.
\label{solasyrun}
\end{eqnarray}
Consequently,   the
asymptotic behavior of the gluon distribution is given by
\begin{eqnarray}
xG(x,Q^2) = \frac{\epsilon}{1+ \epsilon} \,\frac{2N_c Q^2 R^2}{3 \pi^2} \,\,
ln \,(\frac{1}{x}) \,\,.  \label{gluonasyrun}
\end{eqnarray}
As expected at large values of $\epsilon$ $(Q^2)$ this solution reduces to
the solution at fixed $\alpha_s$. The main difference is the behavior at
small values of $\epsilon$, where the prefactor in (\ref{gluonasyrun}) is
important. However, the partial saturation of the gluon distribution is not
modified by the running of $\alpha_s$. It agrees with the result that the
unitarity corrections are expected to be relevant before  the next to
leading order corrections \cite{mueplb}.

Before we present our results and conclusions in the next section, let us discuss the
generality of the above results for the gluon distribution by the comparison with the
predictions of  other high density approach. In contrast to the AGL formalism, the
approach proposed by McLerran, Venugopalan and collaborators (MV-JKLW approach)
\cite{mv,jamal} address the high density regime from the nonperturbative region by
developing an effective Lagrangian approach to a very dense system.
 In Ref. \cite{jamal2}
the authors have considered the DLA limit of the functional evolution equation 
derived from this approach,  obtaining 
 an  equation similar, but not  identical, to the AGL equation (we refer
\cite{jamal2} for details). One of
the main results of this work  was the solution of the proposed equation for  large values of densities
(large $\kappa$), which is given by
\begin{eqnarray}
xG(x,Q^2) = \frac{N_c(N_c - 1) \pi}{2}\,R^2 \,Q^2\, ln \left(\frac{1}{x}%
\right)\,\,.  \label{mvdlasol}
\end{eqnarray}
The remarkable feature of this   solution is  the same $Q^2$ and $x$ dependence of the
AGL equation in the asymptotic regime. The difference in the prefactors is a function
of the distinct normalizations and approximations used in the two approaches.

\section{The Asymptotic Behavior of $F_2$}

The main conclusion of our analysis in the previous section was the universality of the
$x$ and $Q^2$ dependence of the gluon distribution (the partial saturation) in the
asymptotic regime of the high density QCD. In this section we will consider the
consequences of the partial saturation in the behavior of the structure function, which
is experimentally measured.

Using the solution of the AGL equation in the asymptotic regime [Eq. (\ref {gluonasy})]
as input in the Eq. (\ref{f2vio}) we get
\begin{eqnarray}
F_2(x,Q^2) \approx \frac{ \alpha_s}{ \pi^3}\,R^2\,Q^2 \,ln \left(\frac{1}{x}%
\right) \,\,.  \label{f2asy}
\end{eqnarray}
We see that the partial saturation of the gluon distribution implies that the $F_2$
structure function does not saturate at small values of $x$, but is linearly
proportional to $ln \, s$. Basically, we verify that the AGL approach, as well as the
MV-JKLW approach maximizes the prediction  for the $F_2$ structure function in the
asymptotic regime of very high densities, resulting in the   black disc prediction,
when the leading twist relation [Eq. (\ref{f2vio})] is used.

We now consider that the relation between the structure function and the gluon
distribution is given by the Eq. (\ref{diseik2}). In the asymptotic regime (large
$\kappa_q$) we obtain
\begin{eqnarray}
F_2(x,Q^2) \approx \frac{R^2}{2\pi^2} \sum_{u,d,s} \epsilon_i^2 \int_{\frac{1%
}{Q^2}}^{\frac{1}{Q_0^2}} \frac{d^2r_t}{\pi r_t^4} \,ln(\kappa_q(x, r_t^2))
\,\,.  \label{diseikasy}
\end{eqnarray}
Using the asymptotic solution of the AGL equation we can determine $\kappa_q = 4/9
\kappa_G$ at large values of densities, and so
\begin{eqnarray}
F_2(x,Q^2) \approx \frac{R^2}{2\pi^2} \sum_{u,d,s} \epsilon_i^2 \int_{\frac{1
}{Q^2}}^{\frac{1}{Q_0^2}} \frac{d^2r_t}{\pi r_t^4} \,ln\,\left[ \frac{4 \alpha_s}{3} \,
ln \, (\frac{1}{x})\right] \,\,.  \label{diseikasy2}
\end{eqnarray}
Therefore, considering the contribution of the higher twist terms in the
relation between the structure function and the gluon distribution, we
predict the following asymptotic behavior for the structure function
\begin{eqnarray}
F_2(x,Q^2) \approx \frac{R^2 Q^2}{3 \pi^2} \,ln\,\left[ \frac{4 \alpha_s}{3}
\, ln \, (\frac{1}{x})\right] \,\,.  \label{diseikasy3}
\end{eqnarray}
We see that the inclusion of the higher twist terms implies a softer dependence of $F_2$
with the energy than obtained using the leading twist relation. In contrast with the
leading twist prediction, in this case the black disc limit is not maximized.  However,
in both cases the structure function does not violate the Froissart boundary in the
asymptotic regime of high density QCD. The demonstration of the behavior
(\ref{diseikasy3}) using the other approaches for hdQCD is  an important open question.
In particular, the analyzes of the predictions for $F_2$  obtained in Ref. \cite{mv2}
in the infinite momentum frame for the very high density regime would be a important
check of our results. Our study demonstrates the importance of the correct connection (leading or higher twist relation)
between the gluon distribution and the observables when we are analyzing the interface
between perturbative and nonperturbative QCD.

 From the results for the structure function in
the asymptotic regime we can see that this regime is characterized by the identity
\begin{eqnarray}
\frac{d F_2(x,Q^2)}{ d ln \, Q^2} = F_2 (x,Q^2)\,\,,
\end{eqnarray}
which is an important signature of the asymptotic regime of high density
QCD. This regime should be reached for the case of an interaction with
nuclei at smaller parton densities than in a nucleon, since $\kappa_A = A^{%
\frac{1}{3}} \times \kappa_N$, where $\kappa_N$ is given by the expression (%
\ref{kapag}).

\section{Conclusions}

In this paper we have analyzed the description of the $F_2$ structure function in the
regime of very high densities, assuming first that the leading twist relation between
$F_2$ and $xG$ is valid and second that this relation is modified by the higher twist
terms associated to the high density  corrections. In the first case we have obtained
that the corresponding $F_2$ structure function is linearly proportional to $Q^2 R^2
\,ln \, 1/x$, which agrees with the results predicted in the black disc limit. This
result is not high density approach dependent. In the second case a softer behavior is
obtained. In both cases, the $F_2$ structure function unitarizes and, as expected by
construction, the black disc boundary is not violated in the asymptotic regime of high
density QCD. We verify that when the parton density is such that the proton becomes
black and the interaction probability is unity, the structure function becomes
proportional to $Q^2$ and a soft energy dependence is predicted. As a by product, we have
derived the AGL equation for running $\alpha_s$ and obtained a signature for the
asymptotic regime of QCD. 

Our main conclusion is that we should be careful to relate
the gluon distribution with the observables (for instance, $F_2$, $F_2^c$, heavy quark
production, ...) in the high density limit, since besides the behavior of the gluon
distribution, also the relation between these quantities must be modified from the
usual leading twist expressions.

\section*{Acknowledgments}

VPBG thanks FAPERGS and CNPq for support. 
This work was partially financed by CNPq and by Programa de Apoio a
N\'ucleos de Excel\^encia (PRONEX), BRAZIL.

\end{document}